# Relatives in the same university faculty: nepotism or merit?[1]

*Giovanni Abramo[a,b,\*], Ciriaco Andrea D'Angelo[b], Francesco Rosati[b]*

[a] Institute for System Analysis and Computer Science (IASI-CNR)
National Research Council of Italy

[b] Laboratory for Studies of Research and Technology Transfer
School of Engineering, Department of Management
University of Rome "Tor Vergata"

**Abstract**

In many countries culture, practice or regulations inhibit the co-presence of relatives within the university faculty. We test the legitimacy of such attitudes and provisions, investigating the phenomenon of nepotism in Italy, a nation with high rates of favoritism. We compare the individual research performance of "children" who have "parents" in the same university against that of the "non-children" with the same academic rank and seniority, in the same field. The results show non-significant differences in performance. Analyses of career advancement show that children's research performance is on average superior to that of their colleagues who did not advance. The study's findings do not rule out the existence of nepotism, which has been actually recorded in a low percentage of cases, but do not prove either the most serious presumed consequences of nepotism, namely that relatives who are poor performers are getting ahead of non-relatives who are better performers. In light of these results, many attitudes and norms concerning parental ties in academia should be reconsidered.

\* *Corresponding author*

# 1. Introduction

Nations with high levels of corruption and higher education systems with no or low intensity of competition among universities are generally more exposed to phenomena of favoritism in faculty recruitment and career advancement. Nepotism is a particular form of favoritism. The circumstances of the phenomenon in academic spheres have been little studied, apart from modest efforts in Italy. The only scientific work we have found on this question in other nations concerns discrimination against career advancement of U.S. women through anti-nepotism regulations, dating to almost 50 years ago (Simon et al., 1966). Other studies on recruitment and career advancement of faculty concern the more general phenomenon of favoritism in Turkey (Aydogan, 2012) and in Australia (Martin, 2009). In Italy (51st in 2011 Corruption Rank[2]) the public sector is notorious for widespread favoritism, meaning that parental links within universities, which are primarily public, are automatically labeled as an expression of nepotism. However, there is no robust empirical verification of the accusation.

There are many nations where parental links at faculty level within universities are discouraged, either by culture or practice. Italy has recently gone so far as to legislate against the phenomenon. Article 18, Law 240 of 30/12/2010 clearly limits who can participate in competitions for associate or full professorships, for research funding or for award of any form of contract issued by the university. Participation is explicitly forbidden to anyone related, up to and including the fourth degree, with any professor belonging to the department or unit that initiates such a competition, or similarly with the rector, director-general, or any member of the university board of trustees.

Allesina (2011) applied standard statistic techniques, based on shared last names among professors, to the set of all 61,340 2010 Italian academics. On the basis of the very simple axiom *high levels of corruption + high homonymity rates = nepotism*, he concludes that "nepotism is prominent in Italy, with particular disciplinary sectors being detected as especially problematic. Out of 28 disciplines, 9 – accounting for more than half of Italian professors – display a significant paucity of different last names". Durante et al. (2009) also outright discard the possibility that the very high homonymity rates in Italian universities could be generated by a random process. The authors link homonymity rates to academic performance by faculty, showing a strong negative relation between the two. However this conclusion is of little interest, given two weaknesses that must not be ignored. The first is that the number of professors who are potential beneficiaries of nepotism is in general quite modest compared to the overall staff, making it difficult to imagine that their performance could determine that of the entire faculty in such a notable manner. The second is that the performance rankings of faculties used by the authors (data from the annual survey on Italian universities published by CENSIS, and from a study by the Conference of Italian University Rectors), derive from a measurement procedure that is too approximate to be considered trustworthy (Abramo et al., 2008). Such performance rankings have in fact never been published in an international journal. In a subsequent work, Durante et al. (2011) repeat their observations using the results of the first national evaluation exercise, but this national exercise again lacked robustness and accuracy, as demonstrated by Abramo and D'Angelo (2011).

On the other hand, many sociological studies show that parents can pass along a substantial amount of career-relevant knowledge to their children (Dunn and Holtz Eakin, 2000; Lentz and Laband, 1989). This phenomenon, rather than nepotism, could contribute to explaining the high homonymity rates in Italian universities.

---

[2] Retrieved from: http://www.worldaudit.org/corruption.htm

The study we propose here overcomes the preceding limits, through direct comparison of research performance of the potential beneficiaries of nepotism against all their Italian university colleagues in the same research field, with the same academic rank and seniority (i.e. number of years with the university). The results contradict the preceding articles and conventional wisdom: even in the context of the widely-diffused practices of favoritism seen in the Italian university system, we do not observe significant differences in performance for the "children" with relations in the same university in comparison to the other professors.

In the next section we outline the distinctive features of the Italian higher education system. In section three we present the methodology applied to measure productivity rankings at the individual level. In section four we analyze the family relationships in our dataset. In section five we present the comparison of research performance of the potential beneficiaries of nepotism against all their Italian university colleagues. Section six presents our concluding remarks.

## 2. The Italian higher education system

The Italian Ministry of Education, Universities and Research (MIUR) officially recognizes a total of 95 universities, with authority to issue legally-recognized degrees. Twenty-eight of these are very small private special-focus universities, of which 12 offer only e-learning. Sixty-seven are public and generally multi-disciplinary universities, scattered throughout the nation, some having a number of branches in smaller towns. Six of them are *Scuole Superiori* (Schools for Higher Studies), specifically devoted to highly talented students, with very small faculties and tightly limited enrollment numbers per degree program. In Italy, 94.9% of faculty are employed in public universities (0.5% in Scuole Superiori) and 5.1% are in private universities.

The Italian higher education system is a long-standing classic example of a public and highly centralized governance structure, with low levels of autonomy at the university level and a very strong role played by the central state[3]. To date, the most significant intervention for liberalization has been Law 168 of 1989, intended to grant increased autonomy and responsibilities to the universities. In particular, Articles 6 and 7 of this law are intended to enact Article 37 of the national Constitution, by establishing the fundamentals of university autonomy in teaching, research, financing and accounting, and directing that individual institutions appropriately establish an autonomous organizational framework including their own charters and regulations. Law 537 (Article 5) of 1993 and Decree 168 of 1996 introduced substantial changes in central financing for universities, specifically in relation to overall amounts, re-equilibration among institutions, freedom and responsibility in allocating expenses, involvement in fund-raising, and freedom to apply tuition fees provided that these do not exceed 20% of total government funding. Law 537 had two objectives: to increase university involvement in overall decision-making relating to use of resources, and to encourage individual institutions to operate on the market and reach their own economic and financial equilibrium. In keeping with the Humboldt model, there are no "teaching-only" universities in Italy, as all professors are required to carry out both research and teaching. National regulations establish that each faculty member must allocate a minimum of 350 hours per year to teaching. At the close of 2010, there were 59,000 faculty members in Italy (full, associate and assistant professors) and a roughly equal number of technical-administrative staff. All new personnel enter the university system through public examinations and career advancement can only proceed by further public examinations. Salaries are regulated at the centralized level and are calculated according to role (administrative, technical, or professorial), rank within role (for example: assistant, associate or full professor) and seniority. None of a professor's salary depends on merit: salaries increase annually according to rules set by government. Moreover, as in all Italian public administration, dismissal of an employee for lack of productivity is unheard of.

---

[3] An in-depth analysis of the Italian higher education system may be found in Boffo et al., 2006

The whole of these conditions create an environment and a culture that are completely non-competitive, yet flourishing with favoritism and other opportunistic behaviors that are dysfunctional to the social and economic roles of the higher education system. The overall result is a system of universities that are almost completely undifferentiated for quality and prestige, with the exception of the tiny Scuole Superiori and a very small number of the private special-focus universities. The system is thus unable to attract significant foreign faculty or students. The numbers are negligible: foreign students are 3% of the total, compared to the OECD average of 8.5%, and only 2.3% of actual graduates are foreigners; only 1.8% of research staff are foreign nationals. This is a system where every university has some share of top scientists, flanked by another share of absolute non-producers. Over the 2004-2008 period, 6,640 (16.8%) of the 39,512 hard sciences professors did not publish any scientific articles in the journals indexed by the Thomson Reuters' *Web of Science* (WoS). Another 3,070 professors (7.8%) did achieve publication, but their work was never cited. This means that 9,710 individuals (24.6%) had no impact on scientific progress. It is not surprising then that no Italian universities is ranked above 150 position in any of the yearly world universities rankings.

**3. Productivity rankings at individual level - methodology**

Research activity is a production process in which the inputs consist of human, tangible (scientific instruments, materials, etc.) and intangible (accumulated knowledge, social networks, etc.) resources, and where outputs have a complex character of both tangible nature (publications, patents, conference presentations, databases, protocols, etc.) and intangible nature (tacit knowledge, consulting activity, etc.). The new-knowledge production function has therefore a multi-input and multi-output character. The principal efficiency indicator of any production system is labor productivity. To calculate it one needs adopt a few simplifications and assumptions. It has been shown (Moed, 2005) that in the hard sciences, including life sciences, the prevalent form of codification of research output is the publication in scientific journals. As a proxy of total output in this work we consider only publications (articles, article reviews, and proceeding papers) indexed in the WoS. The other forms of output which we neglect are often followed by publications that describe their content in the scientific arena, so the analysis of publications alone actually avoids a potential double counting.

When measuring labor productivity, if there are differences in the production factors available to each scientist then one should normalize by them. Unfortunately relevant data are not available at individual level in Italy. The first assumption then is that resources available to professors within the same field of observation are the same. The second assumption is that the hours devoted to research are more or less the same for all professors. In Italy the above assumptions are acceptable, because in the period of observation core government funding was input oriented, and distributed to satisfy the resource needs of each and every university in function of their size and activities. Furthermore, the hours that each professor has to devote to teaching are established by national regulations and the same for all.

Research projects frequently involve a team of researchers, which shows in co-authorship of publications. Productivity measures then need to account for the fractional contributions of scientists to their outputs. In the life science, the position of co-authors in the list reflects the relative contribution to the project and needs to be weighted accordingly. Furthermore, because the intensity of publications varies across fields (Abramo et al., 2008), in order to avoid distortions in productivity rankings, one must compare researchers within the same field. A prerequisite of any research performance assessment free of distortions is then a classification of each researcher in one and only one field. In the Italian university system all professors are classified in one field. To our knowledge, this feature of the Italian higher education system is unique in the world. In the hard sciences, there are 205 such fields (named scientific disciplinary sectors, $SDSs^4$), grouped into nine disciplines (named university disciplinary areas, $UDAs^5$). Since it has been demonstrated that productivity of full, associate and assistant professors is different (Abramo et al., 2011), and academic rank determines differentiation in stipends, comparisons of research performance should be differentiated by academic rank.

A very gross way to calculate the average yearly labor research productivity is to simply measure the weighted fractional count of publications per researcher in the period of observation and divide it for the full time equivalent of work in the period. A more sophisticated way to calculate productivity recognizes the fact that publications, embedding the new knowledge produced, have different values. Their value depends on their impact on scientific advancements. As proxy of impact bibliometricians adopt the number of citations for the researchers' publications.

However, comparing researchers' performance by field and academic rank is not enough to avoid distortions in rankings. In fact citation behavior too varies across fields, and it has been shown that it is not unlikely that researchers belonging to a particular scientific field may also publish outside that field (a typical example is statisticians, who may apply theory to medicine, physics, social sciences, etc.). For this reason we standardize the citations for each publication accumulated at June 30, 2009 with respect to the median[6] for the distribution of citations for all the Italian publications of the same

---

[4] The complete list is accessible on http://attiministeriali.miur.it/UserFiles/115.htm

[5] Mathematics and computer sciences; physics; chemistry; earth sciences; biology; medicine; agricultural and veterinary sciences; civil engineering; industrial and information engineering.

[6] As frequently observed in literature (Lundberg, 2007), standardization of citations with respect to median value rather than to the average is justified by the fact that distribution of citations is highly skewed in almost all disciplines.

year and the same subject category[7].

In formulae, the average yearly productivity at the individual level, $P$ is the following:

$$P = \frac{1}{t} \cdot \sum_{i=1}^{N} \frac{c_i}{Me_i} * \frac{1}{s_i}$$

Where:
$t$ = number of years of work of the researcher in the period of observation
$c_i$ = citations received by publication $i$;
$Me_i$ = median of the distribution of citations received for all Italian cited-only publications of the same year and subject category of publication $i$;
N = number of publications of the researcher in the period of observation.
$s_i$ = co-authors of publication $i$

In the life sciences, widespread practice is for the authors to indicate the various contributions to the published research by the positioning of the names in the authors list. For life sciences then, when the number of co-authors is higher than two, different weights are given to each co-author according to his/her position in the list and the character of the co-authorship (intra-mural or extra-mural). If first and last authors belong to the same university, 40% of citations are attributed to each of them; the remaining 20% are divided among all other authors. If the first two and last two authors belong to different universities, 30% of citations are attributed to first and last authors; 15% of citations are attributed to second and last author but one; the remaining 10% are divided among all others[8].

Based on the value of $P$ we obtain, for each SDS, a ranking list expressed in percentiles and differentiated by academic rank. Thus the performance of each scientist is calculated in each SDS for each academic rank and expressed on a percentile scale of 0-100 (worst to best) for comparison with the performance of all Italian colleagues of the same academic rank and SDS. We can exclude, for the Italian case, that productivity ranking lists may be distorted by variable returns to scale, due to different sizes of universities (Abramo et al., 2012).

**4. Family relationships**

The starting point for identifying family links within the same university is the identification of professors with the same last name (Allesina, 2011; Durante et al., 2009; Durante et al., 2011; Angelucci et al., 2010; Güell et al., 2007). Pairs are then identified among the homonymous professors. For convenience we label the pairs as "parent-child", even if they could be grandparent-grandchild, uncle/aunt-nephew/niece, brother-sister, cousins, or unrelated. This procedure based on the same last names, inevitably excludes identification of most family relations headed by the "mother", and wife-husband relations. To make the identification of professors potentially subject to nepotism more reliable we impose the following conditions. The field of observation from which "children" are extracted concerns assistant and associate professors who joined a faculty or advanced rank in the years 2001, 2002 and 2003, totaling 13,607; while the "parents" are the full professors of the years of 2000[9], 2001, 2002 and 2003, totaling 13,598. Further the family name must not be included on the list of the 500 most common surnames in Italy[10], nor among the 20 most common surnames in the region where the university is based[11].

---

[7] The subject category of a publication corresponds to that of the journal where it is published. For publications in multidisciplinary journals the scaling factor is calculated as a weighted average of the standardized values for each subject category.
[8] The weighting values were assigned following advice from Italian professors in the life sciences. The values could be changed to suit different practices in other national contexts.

The Italian Ministry of Education, University and Research (MIUR), maintains a database (http://cercauniversita.cineca.it), showing for each Italian professor: full name, university, department, SDS, and academic rank.

We identified 860 family links corresponding to the above criteria, with 28.7% of the relationships involving a parent and child in the same UDA. In 10.3% of cases, the parent and child belong to the same SDS. Some links involve more than two subjects (292, or 34.0% of the total links), specifically:

- One child – many parents (131, equal to 15.2% of the total links): in this case the situation is identified as a single parent-child pair. Among these, in 21 cases (2.4%) the number of parents is equal to or greater than three.
- Many children – one parent (112, equal to 13.0% of the total links): the number of pairs is equal to the number of children. Among these, in 16 cases (1.9%) the number of children is equal to or greater than three.
- Many children – many parents (49, or 5.7% of the total links): the number of pairs is equal to the number of children. Among these, in 9 cases (1.0%) both the number of children and the number of parents are equal to or greater than three.

Of the 860 possible original links, 764 are defined in the above terms. The "children" compose 5.6% of the 13,607 academics of the overall dataset. The analysis of the distribution of links per UDA shows the highest concentration (7.1%) in the Medicine UDA, and the lowest in the UDA for the Antiquities, philological-literary and art-historic sciences. The distribution of pairs by geographic area of university location shows the maximum concentration in southern Italy (379 children, equal to 7.7% of the 4,924 "potential" children, in universities situated in the south), while the minimum concentration (3.8%) is in the north. This could be due to two causes: one is the migratory flux from the poorer south to the richer north, which has always characterized the nation, and the other is the possibly greater diffusion of the phenomenon of nepotism in the south (Allesina, 2011; Durante et al., 2011). The higher concentration of homonymy in the south could also be aggravated by the inclusion of Italy's two large islands.

## 5. Research performance

### 5.1 Research performance of children and non-children

To ensure the representativity of publications as proxy of the research output, the field of analysis is limited to those SDSs where at least 50% of professors produced at least one publication in the period 2004-2008. The calculations of performance also exclude all professors who were not on faculty for at least three years between 2004 and 2008. With these restrictions, the evaluation is limited to 193 SDSs, 36,931 professors and 493 children (thus also 493 pairs), with 443 parents.

---

[9] To qualify as potential nepotism for children selected to faculty in 2001, we impose the condition that the "parent" must have been a full professors at least since the preceding year.
[10] Retrieved from the "International Laboratory of Onomastics".
http://onomalab.uniroma2.it/contents/allegati/3000_cognomi_italia_2000.pdf. Last accessed on January 20, 2014.
[11] Retrieved from http://www.cognomix.it/. Last accessed on January 20, 2014.

We then compared the performance of children with non-children of the same SDS and the same academic rank to the end of 2008 (Table 1). The average percentile rank of the children is 55.9, compared to 54.3 for non-children with the same seniority, and 45.1 for all non-children. The Student's *t*-test for significance in the differences gave negative results (two tailed *P*-value = 0.301) for the first two rankings and positive (two tailed *P*-value <0.001) for the comparison between children and all non-children. The share of children with no publications is 6.9% of total, compared to 9.7% of the non-children with the same seniority and 18.4% of all non-children; the children with no citations are 15.2%, compared to 15.6% of the non-children with the same seniority and 25.1% of all non-children. From this point we limited the comparisons to the performance of children against that of non-children with the same seniority.

The comparison between research performance of children and of non-children, by UDA (Table 2), does not show significant differences, according to Student's *t*-test conducted for each UDA.

*Table 1: Comparison between 2004-2008 research performance of children and of non-children*

|  | Children | Non-children (same seniority) | All Non-children |
|---:|:---:|:---:|:---:|
| Observations | 493 | 6,723 | 21,542 |
| Average percentile rank | 55.9 | 54.3 | 45.1 |
| Professors with no publications (%) | 6.9 | 9.7 | 18.4 |
| Professors with no citations (%) | 15.2 | 15.6 | 25.1 |
| Above median (%) | 61.1 | 59.9 | 48.2 |
| Top 20% scientists (%) | 30.0 | 26.5 | 19.5 |
| Top 10% scientists (%) | 15.6 | 14.2 | 9.9 |
| Absolute top scientists (%) | 2.6 | 1.9 | 1.2 |

*Table 2: Comparison of 2004-2008 research performance of children (C) and of non-children (NC) by UDA*

| UDA | Observations | | Average percentile rank | | Professors with no publications (%) | | Professors with no citations (%) | | Top 20% scientists (%) | | Top 10% scientists (%) | | Absolute top scientists (%) | |
|---|---:|---:|---:|---:|---:|---:|---:|---:|---:|---:|---:|---:|---:|---:|
| | C | NC | C | NC | C | NC | C | NC | C | NC | C | NC | C | NC |
| Mathematics and computer science | 27 | 469 | 55 | 59.1 | 3.7 | 3.6 | 22.2 | 13.4 | 29.6 | 30.1 | 18.5 | 15.8 | 3.7 | 1.1 |
| Physics | 17 | 382 | 56.1 | 61.7 | 5.9 | 0.8 | 5.9 | 2.4 | 17.6 | 30.4 | 5.9 | 15.2 | 5.9 | 1 |
| Chemistry | 41 | 526 | 58 | 58.5 | 0 | 0.6 | 2.4 | 0.8 | 31.7 | 27.6 | 17.1 | 13.7 | 0 | 1.3 |
| Earth sciences | 14 | 178 | 46.6 | 52.4 | 28.6 | 14.6 | 28.6 | 21.3 | 21.4 | 24.7 | 7.1 | 15.2 | 0 | 2.8 |
| Biology | 61 | 910 | 57.9 | 55.4 | 8.2 | 5.9 | 8.2 | 8.9 | 31.1 | 26.2 | 18 | 13.2 | 3.3 | 1.1 |
| Medicine | 185 | 2,324 | 56.4 | 53 | 7.6 | 13.8 | 14.6 | 18.2 | 28.1 | 25.1 | 13.5 | 13.8 | 2.7 | 1.6 |
| Agricultural and veterinary sciences | 45 | 527 | 48.6 | 51.1 | 4.4 | 10.2 | 24.4 | 20.1 | 24.4 | 24.7 | 15.6 | 14.6 | 2.2 | 3.8 |
| Civil engineering and architecture | 15 | 253 | 51.7 | 51.9 | 6.7 | 11.9 | 33.3 | 28.1 | 26.7 | 27.7 | 6.7 | 13.4 | 0 | 0.8 |
| Industrial and information engineering | 69 | 829 | 58.2 | 55.7 | 5.8 | 6.4 | 15.9 | 14.7 | 40.6 | 28.5 | 21.7 | 15.1 | 4.3 | 2.9 |
| Other | 19 | 325 | 59.6 | 43.1 | 10.6 | 27.4 | 21.7 | 39 | 36.1 | 22.4 | 20.6 | 13.3 | 0 | 3.4 |
| *Total* | *493* | *6,723* | *55.9* | *54.3* | *6.9* | *9.7* | *15.4* | *15.6* | *30* | *26.5* | *15.6* | *14.2* | *2.6* | *1.9* |

From the analysis of performance by geographic area (Table 3), we observe that children are more productive than non-children in every area, but the differences are statistically significant under Student's *t*-test only for Central Italy (two tailed *P*-value = 0.041).

*Table 3: Comparison between 2004-2008 research performance of children (C) and of non-children (NC) by geographic area*

|  |  | North | Centre | South |
|---|---|---|---|---|
| Observations | C | 135 (27.4%) | 125 (25.4%) | 233 (47.3%) |
|  | NC | 2,444 (36.4%) | 1,901 (28.3%) | 2,378 (35.4%) |
| Average percentile rank | C | 62.7 | 58.1 | 50.7 |
|  | NC | 60.0 | 52.0 | 50.3 |
| Professors with no publications (%) | C | 4.4 | 7.2 | 8.2 |
|  | NC | 6.7 | 12.6 | 10.6 |
| Professors with no citations (%) | C | 8.9 | 14.4 | 19.3 |
|  | NC | 11.6 | 18.4 | 17.7 |
| Above median (%) | C | 66.7 | 65.6 | 55.4 |
|  | NC | 66.9 | 57.8 | 54.5 |
| Top 20% scientists (%) | C | 39.3 | 30.4 | 24.5 |
|  | NC | 34.0 | 23.9 | 20.8 |
| Top 10% scientists (%) | C | 23.7 | 16.0 | 10.7 |
|  | NC | 19.6 | 12.7 | 9.7 |
| Absolute top scientists (%) | C | 3.7 | 4.8 | 0.9 |
|  | NC | 2.7 | 1.6 | 1.1 |

## 5.2 Analysis of career advancement

Nepotism in career advancement includes the promotion of children regardless of their research performance, or from a group of candidates of equal performance, or of better performance. In operational terms, to make our conclusions as robust as possible, we test the nepotism hypothesis on a very low performance threshold: if performance of children is within the bottom 20%, we regard their career advancement as potential cases of nepotism.

We then divided the children and non-children of the dataset in two subgroups, distinguishing those who advanced in academic rank from all others (Table 4). We see that 20.9% of the children advanced their career rank, similar to the 20.6% of non-children. In general, promotion of children seems to have been justified by higher performance levels (68.6 average rank, against respectively 52.5 and 51.4 averages for children and non-children who did not advance in rank).

From Student's *t*-test, we see that the differences are statistically significant for the comparison between children who advanced and non-children who did not advance (two tailed *P*-value < 0.001), and also in the comparison between children who did not advance and non-children who advanced in rank (two tailed *P*-value < 0.001).

However we do observe potential cases of nepotism, being 7.8% of the bottom 20% of children advancing in rank; and a still higher incidence of favoritism, being 10.6% of the bottom 20% of non-children advancing in rank. On the other hand, we also see greater penalization among children: 25.6% of the top 20% of children versus 22.8% of the top 20% non-children do not advance in academic rank.

*Table 4: Analysis of career advancement of children and non-children*

|  | Children | | Non-children | |
|---:|:---:|:---:|:---:|:---:|
|  | No career advancement | Career advancement | No career advancement | Career advancement |
| Observations | 390 (79.1%) | 103 (20.9%) | 5,337 (79.4%) | 1386 (20.6%) |
| Average percentile rank | 52.5 | 68.6 | 51.4 | 65.7 |
| Professors with no publications (%) | 8.7 | 0.0 | 11.2 | 3.8 |
| Professors with no citations (%) | 17.7 | 5.8 | 17.7 | 7.9 |
| Bottom 10% scientists (%) | 17.9 | 5.8 | 18.2 | 8.5 |
| Bottom 20% scientists (%) | 19.2 | 7.8 | 21.4 | 10.6 |
| Above median (%) | 56.7 | 77.7 | 56.4 | 73.4 |
| Top 20% scientists (%) | 25.6 | 46.6 | 22.8 | 40.5 |
| Top 10% scientists (%) | 12.3 | 28.2 | 11.6 | 24.1 |
| Absolute top scientists (%) | 1.3 | 7.8 | 1.3 | 4.1 |

**5.3 Performance comparison of parents with non-parents and with children**

The average performance of parents is worse than that of "non-parents" in the same research fields (Table 5). The subsequent Student's *t*-test shows that the difference is significant (two tailed *P*-value = 0.043). Deeper analysis showed that the difference is essentially due to the low performance of parents who retired during this period.

*Table 5: Comparison between 2004-2008 research performance of parents and of non-parents*

|  | Parents | Non-parents |
|---:|:---:|:---:|
| Observations | 443 | 6,540 |
| Average percentile rank | 42.0 | 45.3 |
| Professors with no publications (%) | 13.5 | 12.8 |
| Professors with no citations (%) | 21.9 | 19.5 |
| Above median (%) | 44.5 | 46.6 |
| Top 20% scientists (%) | 16.3 | 19.4 |
| Top 10% scientists (%) | 8.1 | 10.0 |
| Absolute top scientists (%) | 0.0 | 1.2 |

Table 6 presents the comparison between percentile ranks of performance by children and parents, calculated with respect to their colleagues of the same academic rank and SDS. The number of observations drops to 339 due to excluding all children and parents with an intra-university link where one of the professors works in an SDS that cannot be evaluated by bibliometric technique.

The comparison shows that the performance of children is generally better than that of parents. The subsequent Student's *t*-test confirmed that the average percentile ranks of children are significantly higher than that of parents (two tailed *P*-value < 0.001).

*Table 6: Comparison of children's 2004-2008 research performance to that of parents*

|  | Children | Parents |
|---:|:---:|:---:|
| Observations | 339 | 339 |
| Average percentile rank | 54.9 | 41.1 |
| Professors with no publications (%) | 7.4 | 15.3 |
| Professors with no citations (%) | 17.7 | 26.5 |
| Above median (%) | 60.2 | 42.2 |
| Top 20% scientists (%) | 29.5 | 15.9 |
| Top 10% scientists (%) | 17.7 | 8.3 |
| Absolute top scientists (%) | 2.4 | 0.0 |

# 6. Conclusions

We imagine a student, a donor, an enterprise or the branches of public administration needing to choose among Italian universities, and that the only information they have is the rate of parental links among the faculty. While the literature until now would have oriented the decision-maker towards choosing the university with the lowest rate, our analyses lead to the conclusion that the choice, in general, can be made independently of this factor. In particular, if the choice concerns universities in central Italy, a high rate of parental ties could represent, all else equal, a situation of greater research productivity.

It must be said that higher productivity may be the consequence not only of higher capacity and dedication, but also of favored treatment. Individuals can obtain advantages by being told about opportunities, being schooled in how to get ahead (who to talk to, how to present oneself), being invited to participate in activities and being given access to research funds and inside information. For example, a scientist might be invited to join a research team, with great opportunities, whereas another is not invited. Those who are given opportunities and take advantage of them may, as a result, end up with greater productivity. Someone else might have had the same capacity and motivation but not been given the same opportunities. This means that nepotism can be compatible with equal or greater productivity by relatives: higher productivity can be the consequence of favored treatment outside of appointment and promotion decisions. Higher productivity though may also be the effect of the substantial amount of career-relevant knowledge that parents may have passed along to their children during their education. The study's findings do not rule out the existence of nepotism, which has been actually recorded in a low percentage of cases, but do not prove either the most serious presumed consequences of nepotism, namely that relatives who are poor performers are getting ahead of non-relatives who are better performers.

Law 240 of 30/12/2010, which expressly forbids the hiring of relatives within the same university, is seen as having absolutely no empirical grounds for legitimacy. Rather it should be considered discriminatory, depriving "children" of the liberty to compete for access to the university faculties where they can best contribute, because of the simple "guilt" of having a relative in the institution.

There are many countries where culture, practice, regulations or law discourage or restrict the presence of relatives within the same university faculty. The high intensity of rivalry among universities in competitive higher education system in itself represents a practical antidote to nepotism. If even in a country like Italy, with a non-competitive university system and high rates of favoritism (Zagaria, 2007; Perotti, 2008), the possible recipients of family favoring do not show research performance lower than their peers, then there is a need for reflection on the possibility of eliminating all restrictions and reconsidering the practices and attitudes that inhibit co-presence of relatives in the same university.

Advancement in bibliometric techniques now permits, at least in the hard sciences, the application of accurate and robust systems for measurement of individual scientific performance (Abramo and D'angelo, 2011; Costas et al., 2010) in support of recruitment decisions, periodic evaluations and career advancement. In nations with high rates of favoritism, with non-competitive higher education systems under strong central government regulation, greater resort to such instruments could avoid legislative and regulatory interventions within the universities, which are discriminatory and abusive of the freedoms of the individual, in an a priori sense. At the same time, such techniques would represent a strong deterrent to all practices of favoritism in university recruitment competitions, above all if the results of comparative evaluations through bibliometric instruments were made available to members of the competition juries and the public at large.

It may also worth considering measures which counterbalance the advantages of having an academic relative, namely access to career-relevant knowledge. One possibility is to ensure that academics and aspiring academics have access to mentors. Having access to mentors would, in part, compensate for the advantages of having family members who provide advice, guidance and support. A mentoring scheme, combined with appointment and promotion procedures that minimize conflicts of interest, might help to counteract concerns about nepotism.

Rather than applying tangible and intangible restrictions, the adoption of such objective support instruments for evaluation, together with competitive mechanisms among universities for access to funds, and among single researchers for initial hiring, advancement, funding and performance bonuses, could initiate a "virtuous circle" of continuous improvement in the higher education system of countries with characteristics such as Italy's.